# Compact sorting of optical vortices by means of diffractive transformation optics

GIANLUCA RUFFATO,[1,2*] MICHELE MASSARI,[1,2] AND FILIPPO ROMANATO[1,2,3]

[1]*Department of Physics and Astronomy 'G. Galilei', University of Padova, via Marzolo 8, 35131 Padova, Italy*
[2]*Laboratory for Nanofabrication of Nanodevices, c.so Stati Uniti 4, 35127 Padova, Italy*
[3]*CNR -INFM TASC IOM National Laboratory, S.S. 14 Km 163.5, 34012 Basovizza, Trieste, Italy*
*Corresponding author: gianluca.ruffato@unipd.it*

**The orbital angular momentum (OAM) of light has recently attracted a growing interest as a new degree of freedom in order to increase the information capacity of today's optical networks both for free-space and optical fiber transmission. Here we present our work of design, fabrication and optical characterization of diffractive optical elements for compact OAM-mode division demultiplexing based on optical transformations. Samples have been fabricated with 3D high-resolution electron beam lithography on polymethylmethacrylate (PMMA) resist layer spun over a glass substrate. Their high compactness and efficiency make these optical devices promising for integration into next-generation platforms for OAM-modes processing in telecom applications.**

In the last decades, increasing attention has been devoted to optical beams carrying orbital angular momentum (OAM) of light [1], with applications in a wide range of fields [2]: particle trapping and tweezing [3, 4], phase contrast microscopy [5], stimulated emission depletion (STED) microscopy [6], astronomical coronagraphy [7], quantum-key distribution (QKD) [8] and telecommunications [9, 10]. These beams present an azimuthal angular dependence $\exp(i\ell\varphi)$, being $\varphi$ the azimuthal coordinate and $\ell$ the OAM content per photon in units of $h/2\pi$. The fact that these fields offer a potentially unbounded state space has made them advantageous for increasing the amount of information that can be encoded onto a single-photon. In the telecom field, the exploitation of this novel degree of freedom provides the possibility to enhance information capacity and spectral efficiency of today's networks both in the radio and optical domains. On the other hand, the exploitation of OAM-mode division multiplexing (OAM-MDM) still presents a few crucial points and requires further optimization and tests before commercial applications, especially in the optical and infrared range. Among all, the choice of the (de)multiplexing method for OAM sorting represents an important task in the design of optical platforms for OAM beam processing. Different solutions have been presented and described in order to sort a set of multiplexed beams differing in their OAM content: interferometric methods [11], optical transformations [12-19], time-division technique [20], integrated silicon photonics [21], coherent detection [22], OAM-mode analyzers [23-25]. Phase-only diffractive optical elements (DOE) appear as the most suitable choice for the realization of passive and lossless, compact and cheap optical devices for integrated (de)multiplexing applications [26]. Diffractive OAM-mode analyzers have been widely presented and exploited in literature for the analysis of an input superposition of optical vortices. Basically, they are calculated as linear superposition of many fork-holograms differing in OAM singularity and carrier spatial frequency [23-25]. The far-field of such optical elements exhibits bright peaks at prescribed positions, whose intensity is proportional to the energy carried by the corresponding OAM channel in the incident beam. On the other hand, the efficiency is inversely proportional to the number of states being sampled, which makes this solution nonviable especially for quantum applications. Another technique which is suitable to be fabricated in a diffractive form is represented by transformation optics. Solutions based on this technique demonstrated how OAM states can be efficiently converted into transverse momentum states through a *log-pol* optical transformation [12]. This can be realized by using a sequence of two optical elements: the un-wrapper and the phase-corrector. The former performs the conformal mapping of a position $(x, y)$ in the input plane to a position $(u, v)$ in the output plane, where $v=a \arctan(y/x)$ and $u= -a \ln(r/b)$, being $r=(x^2+y^2)^{1/2}$, $a$ and $b$ design parameters. Its phase function $\Omega_1$ is as it follows:

$$\Omega_1(x,y) = \frac{2\pi a}{\lambda f_1}\left[ y\arctan\left(\frac{y}{x}\right) - x\ln\left(\frac{\sqrt{x^2+y^2}}{b}\right) + x + \frac{x^2+y^2}{2a}\right]$$

(1)

The two free parameters $a$ and $b$ determine the scaling and position of the transformed beam respectively. The parameter $a$ takes the value $d/2\pi$, ensuring that the azimuthal angle range $(0, 2\pi)$ is mapped onto a length $d$ which is shorter than the full width of the second element. The parameter $b$ is optimized for the particular physical dimensions of the sorter and can be chosen independently. The phase-corrector, placed at a distance $f_1$, has a phase function $\Omega_2$ given by:

$$\Omega_2(u,v) = -\frac{2\pi ab}{\lambda f_1} \exp\left(-\frac{u}{a}\right)\cos\left(\frac{v}{a}\right) + \alpha u + \beta v \quad (2)$$

where $(\alpha, \beta)$ are fixed carrier spatial frequencies which control the position of the signal on the screen in far-field. A lens with focal length $f_2$ is inserted after the phase-corrector element in order to focus the transformed beam onto a specified lateral position, which moves proportionally to the OAM content $\ell$ according to:

$$\Delta s = \frac{f_2 \lambda}{2\pi a}\ell \quad (3)$$

Alternatively, the focusing quadratic term can be integrated in the phase-corrector element as well. The same setup has been demonstrated to work as multiplexer, in reverse [17, 18]. In its first realization, the two elements were implemented with spatial light modulators (SLMs) [12]. At a later stage [13, 14], they were replaced by two freeform refractive optical components, exhibiting higher efficiency, though not a small size. More recently, the two optical elements have been realized in a diffractive version [19]. In this work, we apply this demultiplexing technique for the OAM sorting of perfect vortices and we further improve the miniaturization level by integrating the two components into a single diffractive optical element (Fig. 1).

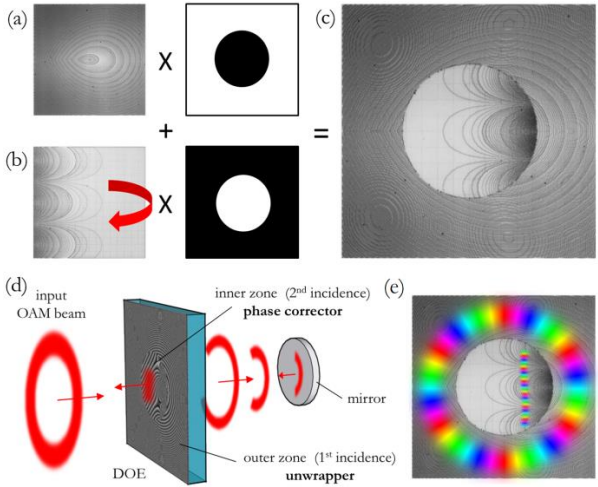

Fig. 1. (a-c) Scheme of the diffractive optical element (DOE) composition. Outer zone: un-wrapper element (a). Inner zone: phase-corrector (b). (d) Scheme of the working principle. (e) Scheme of input and output phase patterns for $\ell=6$.

The stable propagation of OAM modes along ring fibers has highlighted the need to excite and manipulate annular intensity distributions with fixed radius and width, regardless of their OAM content [27, 28]. However, conventionally-generated OAM beams, e.g. Laguerre-Gaussian or Kummer beams, exhibit a ring diameter that increases with the topological charge and therefore they are rather incompatible with applications involving ring fiber cross-section or limited-size optical elements. The so-called perfect vortices appear as the best solution in order to excite optical vortices whose ring dimensions can be adjusted and controlled independently of the OAM content [29, 30]. In our case of interest, the choice of perfect vortices for OAM encoding leaves the central region of the optical element unexploited, since the device acts merely on the zone with non-null incident field. Therefore the non-illuminated central region can be exploited for integrating the second element designed for phase-correction. In this novel configuration, after passing the outer unwrapping zone, the beam is back-reflected by a mirror and goes through the optical element again across the inner central region (Fig. 1.d). Moreover, the optical device is realized into a diffractive form, avoiding bulky refractive elements and allowing a further miniaturization level, especially when short focal lengths are required. The whole DOE phase pattern $\Omega_{DOE}$ is the composition of the two optical elements, according to:

$$\Omega_{DOE} = \Omega_1 \Theta(\rho - \rho_1) + \Omega_2 \Theta(\rho_1 - \rho) \quad (4)$$

being $\rho_1$ the radius of the central zone, $\Theta$ the Heaviside function ($\Theta(x)=1$ for $x>0$, $\Theta(x)=0$ otherwise), provided the condition $\rho_1 > \pi a$ is satisfied (Fig. 1.e). In addition, this solution makes the alignment operations significantly easier, since un-wrapper and phase-corrector result automatically aligned and parallel to each other by design.

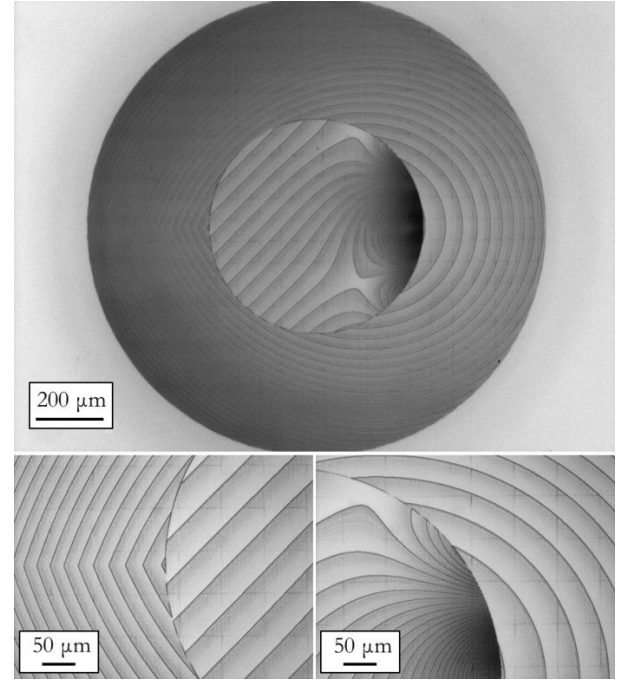

Fig. 2. Sample for integrated transformation optics, optical microscope inspection and details. Design parameters: focusing outer zone, transformation optics values $a=70$ μm, $b=50$ μm, $f_1=20$ mm, carrier spatial frequency in the inner zone for image tilt: $\alpha=\beta=0.1$ μm$^{-1}$. Optimized for wavelength $\lambda=632.8$ nm. 256 phase levels.

We fabricated and characterized the optical response of such a diffractive optical element designed for the demultiplexing of perfect vortices at $\lambda=632.8$ nm. Diffractive optical elements are fabricated as surface-relief patterns of pixels. This 3-D structure can be realized by shaping a layer of transparent material, imposing a direct proportionality between the thickness of the material and the phase delay. Electron beam lithography (EBL) is the ideal technique in order to fabricate 3D profiles with high resolution [31-33]. In this work, the DOE patterns were written on a polymethylmethacrylate (PMMA) resist layer with a JBX-6300FS JEOL EBL machine, 12 MHz, 5 nm resolution, working at 100 keV with a current of 100pA. The substrate used for the fabrication is glass, coated with an ITO layer with conductivity of 8-12 Ω in order to ensure both transparency and a good discharge during the exposure. Patterned samples were developed under slight agitation in a temperature-controlled developer bath for 60 s in a solution of deionized water: isopropyl alcohol (IPA) 3:7. For the considered wavelength of the laser beam used ($\lambda$=632.8nm), PMMA refractive index results $n_{PMMA}$=1.489 from spectroscopic ellipsometry analysis (J.A. Woollam VASE, 0.3 nm spectral resolution, 0.005° angular resolution). The height $d_k$ of the pixel belonging to the $k$th layer is given by:

$$d_k = \frac{k-1}{N} \frac{\lambda}{n_{PMMA}-1} \quad k=1,\ldots,N \quad (5)$$

being $N$ the number of phase levels. In our case of interest, for $N$=256 we get: $d_1$=0 nm, $d_{256}$=1289.0 nm, step $\Delta d$=5.1 nm. In Fig. 2 a DOE sample is shown, with design parameters $f_1$=20 mm, $a$=70 μm, $b$=50 μm. A tilt term was added to the phase corrector with carrier spatial frequencies $\alpha$=$\beta$=0.1 μm$^{-1}$, in order to prevent any overlapping with the potentially noise-carrier zero-order term.

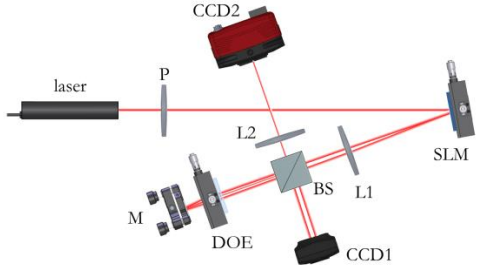

Fig. 3. Scheme of the experimental setup. The laser beam is linearly polarized and illuminates a SLM display for perfect vortex generation. A first lens (L1) collimates the beam which is split with a 50:50 beam-splitter (BS) for OAM beam analysis with the first camera (CCD1). The second part of the beam illuminates the DOE on a sample holder. The optical vortex illuminates the DOE outer zone (un-wrapper) and is reflected back by a mirror (M) upon the DOE inner region (phase-corrector). Then the beam is Fourier-transformed with a lens (L2) collected on the second CCD camera (CCD2).

There is a crucial point to be considered during miniaturization: the OAM-band of the sorting optics. Since the sorting element is designed to work perfectly for collimated beams illuminating the unwrapping term, for non-collimated beams it is required that the angular deviation introduced by the un-wrapper dominates over any deviation from normal incidence. If the input vortex is a ring with radius $R_V$ and the separation between the two elements is $f_1$, the angular deviation introduced by the first element is approximately $R_V/f_1$ for small angles. Irrespective of their divergence, OAM beams are never plane waves, as there is an azimuthal component of the Poynting vector whose local skew angle is around $\ell/(kR_V)$. It follows that for the sorter to work properly it is required that $\ell/(kR_V)<<R_V/f_1$, which imposes [14]:

$$\ell_{MAX} << \frac{2\pi R_V^2}{\lambda f_1} \quad (6)$$

This bound could be increased by either reducing the focal length $f_1$ or increasing the vortex radius $R_V$, i.e. the optical element size. In order to correctly process OAM values up to $\ell\sim20$, while keeping a millimetric size of the optics, we chose a focal length $f_1$=9 mm and an external radius of 1 mm for the unwrapping zone. A Fresnel correction was added to the phase-correcting term. The design parameters are $a$=170 μm, $b$=400 μm, $\alpha$=$\beta$=0.1 μm$^{-1}$. The radius of the inner zone is $\rho_1$=650 μm.

The characterization setup was mounted on an optical table (Fig. 3). The Gaussian beam ($\lambda$= 632.8 nm, beam waist $w_0$=240 μm, power 0.8 mW) emitted by a HeNe laser source (HNR008R, Thorlabs) is linearly polarized before illuminating the display of a LCoS spatial light modulator (PLUTO-NIR-010-A, Holoeye, 1920 x 1080 pixels, 8 μm pixel size, 8-bit depth) for vortex generation. The reflected beam is Fourier-transformed with a first lens of focal length $f_0$=25 cm and a beam-splitter is used to split the beam and analyze field profile and OAM content at the same time. The field profile is collected with a first CCD camera. The beam illuminates the first outer zone of the DOE sample, mounted on a XY translation sample holder with micrometric drives. A mirror is placed on a kinematic mount (KM100, Thorlabs) and the distance is controlled with a micrometric translator (TADC-651, Optosigma). After passing through the DOE inner zone, the signal is collected by a second CCD camera placed at the back-focal plane of a lens with $f_2$=10.0 cm.

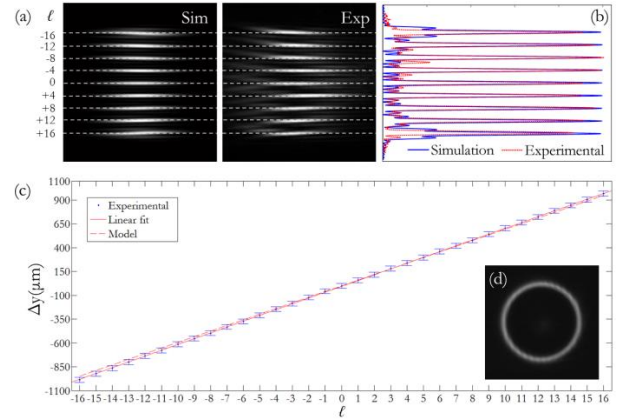

Fig. 4. (a) Simulated and experimental far-field for different input $\ell$ in the range $\ell$={-16,…,+16}, step $\Delta\ell$=4, and cross-sections (b). (c) Position of the experimental spot produced at the plane of the CCD as a function of $\ell$ and linear fit. Error bars: full-width-half-maximum of the intensity peak. Experimental slope: $(\Delta y/\ell)_{exp}$=60.9 μm. Theoretical slope according to Eq. (3): $(\Delta y/\ell)_{th}$=59.3 μm. (d) Measured intensity of the incident perfect vortex with $\ell$=+2, on a plane perpendicular to the propagation direction: ring radius 810 μm, ring width around 67 μm.

The SLM implements a phase mask $\Omega_{SLM}$ combining axicon and spiral phase functions with a quadratic-phase term for curvature correction:

$$\Omega_{SLM}(r,\varphi) = \ell\varphi + \gamma r + k\frac{r^2}{2R} \quad (7)$$

where $\gamma$ is the axicon parameter, $k$ is the wavevector in air, and $R$ is the curvature radius of the incident beam. At the back focal plane of the first lens, a ring is formed having the diameter $R_V$ given by $R_V \simeq \gamma f_0/k$ and topological charge $\ell$. The ring-width $\Delta R_V$ can be controlled by changing the incident beam waist $w$, according to $\Delta R_V = 2f_0/(kw)$. In our case of interest, the ring width results around $\Delta R_V$=65.5 μm. For $\gamma$=0.032μm$^{-1}$, we expect $R_V$=805.7 μm.

The illuminated CCD area is split into rectangular regions which are centered on each elongated spot in far-field with a size corresponding to the minimum separation between any two adjacent channels. We defined 33 of these regions, and we analyzed the optical response under illumination with pure perfect vortices with $\ell$ values in the range from $\ell$ =−16 to $\ell$ =+16 (Fig. 4).

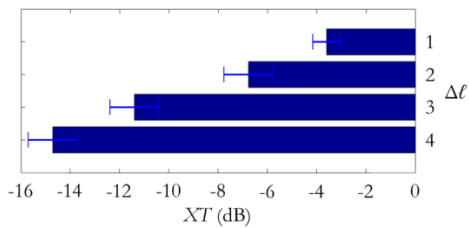

Fig. 5. Average inter-channel cross-talk $XT$ as a function of channel separation $\Delta\ell$ in the range {-16,…,+16}, experimental data.

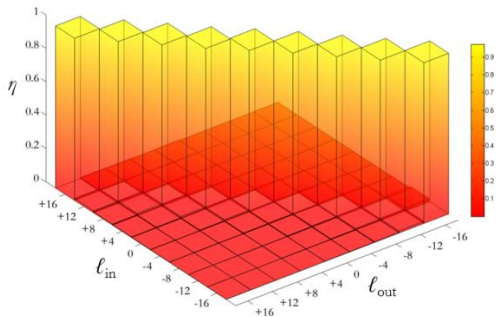

Fig. 6. Total intensities in all detector regions for perfect vortex input modes, experimental data. Input $\ell$ in the range $\ell$={-16,…,+16}, step $\Delta\ell$=4.

As highlighted elsewhere [12-14], a limitation of this sorting technique is represented by the slightly overlap between adjacent channels, which badly affects the inter-channel cross-talk. A solution is to include a fan-out element [15, 16], which increases the phase gradient by introducing multiple copies, thus providing a larger separation between spots, at the expense of increased complexity and size of the optical system. Alternatively, the choice of non-consecutive OAM values further diminishes cross-talk, as illustrated in Fig. 5, where values are reported for increasing channel separation from $\Delta\ell$=1 to 4. The choice of $\Delta\ell$=4 provides a good separation of far-field spots (Fig. 4) and allows obtaining cross-talk values below -15 dB (Fig. 5), with efficiencies up to 96% and off-diagonal terms lower than 2% (Fig. 6).

In conclusion, the fabricated phase masks successfully enable efficient OAM-MDM with a single diffractive optical element performing log-pol optical transformation. The exploitation of perfect vortices is essential in order to control radius and width of the generated ring-shaped beam and, in particular, to make it independent of its OAM value illuminating the outer unwrapping zone of the diffractive optics without overlapping with the central phase-corrector term. In addition, the integration into a single element improves the miniaturization and simplifies alignment operations. The presented optics is promising for integration into optical platforms performing optical processing of OAM channels, for applications both in free-space and in optical fibers.